\documentclass[aps,pra,showpacs,amsmath,amssymb,superscriptaddress,longbibliography,reprint,10pt]{revtex4-1}
\usepackage{graphicx}
\usepackage{xcolor}
\usepackage{url}
\usepackage[normalem]{ulem}
\usepackage{newtxtext,newtxmath}
\usepackage{bm}
\usepackage{bbm}
\usepackage{mathrsfs} 
\usepackage{braket}
\usepackage[colorlinks=true,breaklinks=true,allcolors=blue]{hyperref}

\newcommand\NN{{\mathbbm{N}}}

\newcommand{\mvec}[1]{\boldsymbol #1}

\DeclareMathOperator{\Span}{\mathrm{span}}

\makeatletter
\let\cat@comma@active\@empty
\makeatother
\makeatletter
\AtBeginDocument{\let\LS@rot\@undefined}
\makeatother

\begin{document}

\title{Integrable active atom interferometry}

\author{Michael Kastner} 
\affiliation{National Institute for Theoretical Physics (NITheP), Stellenbosch 7600, South Africa} 
\affiliation{Institute of Theoretical Physics,  University of Stellenbosch, Stellenbosch 7600, South Africa}

\author{Vincent Menet} 
\affiliation{Universit\'e de Lyon, \'Ecole Normale Sup\'erieure de Lyon, 46 All\'ee d'Italie, 69364 Lyon, France}

\author{Johannes N. Kriel} 
\affiliation{Institute of Theoretical Physics,  University of Stellenbosch, Stellenbosch 7600, South Africa}

\date{\today}

\begin{abstract}
Active interferometers are designed to enhance phase sensitivity beyond the standard quantum limit by generating entanglement inside the interferometer. An atomic version of such a device can be constructed by means of a spinor Bose--Einstein condensate with an $F=1$ groundstate manifold in which spin-changing collisions create entangled pairs of $m=\pm1$ atoms. We use Bethe Ansatz techniques to find exact eigenstates and eigenvalues of the Hamiltonian that models such spin-changing collisions. Using these results, we express the interferometer's phase sensitivity, Fisher information, and Hellinger distance in terms of the Bethe rapidities. By evaluating these expressions we study scaling properties and the interferometer's performance under the full Hamiltonian that models the spin-changing collisions, i.e., without the idealising approximations of earlier works that force the model into the framework of SU(1,1) interferometry.
\end{abstract}


\maketitle 

\section{Introduction}
Atom interferometry uses the wave character of atoms, and in particular the superposition principle, to detect phase differences and perform high-precision measurements in a variety of fields, ranging from measurements of the fine structure constant to gravimetry and atomic clocks \cite{CroninSchmiedmayerPritchard09}. The more commonly used passive interferometers use beam splitters that redistribute a conserved number of atoms among two or more modes. Upon recombining the split beams, the phase difference accrued inside the interferometer is measured in the form of interference fringes. The precision to which this phase difference can be measured is specified by the phase sensitivity $\Delta\phi$, which is an important characteristic of any interferometer. The larger the number $N$ of atoms measured at the interferometer's output, the lower can the statistical error be pushed and the higher a phase sensitivity can be reached. Assuming at most classical correlations between the (typically uncorrelated) probed events, passive interferometers are known to have phase sensitivities constrained by the {\em standard quantum limit}, $\Delta\phi\geq1/\sqrt{N}$, which is essentially a consequence of the central limit theorem \cite{GiovannettiLloydMaccone11}. One way to surpass the standard quantum limit is to feed the interferometer with suitably entangled input states. In this case, the {\em Heisenberg limit}\/ $\Delta\phi\geq1/N$, which is a fundamental constraint resulting from Heisenberg's uncertainty principle, may be approached \cite{Caves81}. 

Another strategy to surpass the standard quantum limit goes back to Yurke, McCall, and Klauder \cite{YurkeMcCallKlauder86}, and consists of exchanging passive beam splitters by active components. These active components generate entanglement within the interferometer, and they may have the advantages of being more robust and their experimental realisation being more practical. Originally such active interferometers had been proposed as optical devices, but more recently active atom interferometers have been built \cite{Gross_etal10,SchulzMSc,Linnemann_etal16,Linnemann_etal17,Chen_etal15}, and their improved sensitivity, beyond the standard quantum limit, has been confirmed. One of these experimental realisations is based on effective three-level systems in a Bose--Einstein condensate of $^{87}$Rb atoms, and uses spin-changing collisions as the active component of the interferometer; see Refs.\ \cite{SchulzMSc,Linnemann_etal16,Linnemann_etal17} for details. The Hamiltonian describing such spin-changing collisions, given in Eq.\ \eqref{e:H}, describes the nonlinear interactions between three species of bosons (corresponding to the three levels effectively taking part in the dynamics).

Previous analytical studies of active atom interferometers made use of additional assumptions on the parameters in the spin-changing Hamiltonian \eqref{e:H}, which were chosen such that the relevant time-evolution operators are exponentials of SU(1,1) generators, which leads to significant simplifications. A direct numerical analysis of the full spin-changing Hamiltonian, without additional assumptions, has been reported in Ref.\ \cite{GabbrielliPezzeSmerzi15}. In the present paper we show that an analytic treatment of the full Hamiltonian \eqref{e:H} is possible without any additional assumptions by exploiting Bethe-Ansatz integrability. We use this method to compute exact eigenstates and eigenvalues of the spin-changing Hamiltonian \eqref{e:H}, and for semi-analytic calculations of phase sensitivities and other quantities of interest for several variants of active atom interferometers. By analysing the scaling properties of phase sensitivities we are able to identify parameter regimes in which the standard quantum limit can be surpassed.

\section{Bosonic three-species Hamiltonian\texorpdfstring{\\}{} with spin-changing collisions}
\label{s:Hamiltonian}
A Hamiltonian describing the s-wave scattering between atomic hyperfine states $\ket{F,\kappa}$ with $\kappa\in\{0,\pm1\}$ can be written in second-quantised form as
\begin{multline}\label{e:H_general}
H_\text{gen}=\sum_\alpha\int d^3x\, \Psi_\alpha^\dagger(\mvec{x})(T+V)\Psi_\alpha(\mvec{x})\\
+\sum_{\alpha\beta\gamma\delta}\Omega_{\alpha\beta\gamma\delta}\int d^3x\, \Psi_\alpha^\dagger(\mvec{x})\Psi_\beta^\dagger(\mvec{x})\Psi_\gamma(\mvec{x})\Psi_\delta(\mvec{x}),
\end{multline}
where the field operator $\Psi_\kappa^\dagger(\mvec{x})$ creates the hyperfine state $\ket{F,\kappa}$ at position $\mvec{x}$. Sums in \eqref{e:H_general} are over $\{0,\pm1\}$, the coefficients $\Omega_{\alpha\beta\gamma\delta}$ specify the two-particle interactions between atoms, and $T$ and $V$ denote single-particle kinetic and potential energy terms. Under a number of assumptions \cite{LawPuBigelow98,SchulzMSc}, including a single-mode approximation for ultracold and strongly confined gases in the absence of atom loss, one can eliminate the spatial part of the Hamiltonian and derive an effective Hamiltonian for only the spin part of the atomic states. Adding the effect of microwave dressing on the atoms \cite{ScullyZubairy,SchulzMSc}, one obtains a Hamiltonian of the form
\begin{multline}\label{e:H}
\hspace{-3mm}H=2\lambda\left[a_0^\dagger a_0^{\phantom{\dagger}} \bigl(a_-^\dagger a_-^{\phantom{\dagger}} + a_+^\dagger a_+^{\phantom{\dagger}}\bigr) + a_0^{\phantom{\dagger}} a_0^{\phantom{\dagger}} a_-^\dagger a_+^\dagger + a_0^\dagger a_0^\dagger a_-^{\phantom{\dagger}} a_+^{\phantom{\dagger}}\right]\\
+(q-\lambda)\bigl(a_-^\dagger a_-^{\phantom{\dagger}} + a_+^\dagger a_+^{\phantom{\dagger}}\bigr)
\end{multline}
where $a_0^\dagger$ and $a_\pm^\dagger$ are creation operators of (the spin part of) the three different hyperfine states, $\lambda$ is a parameter related to the scattering strength, and $q$ quantifies the microwave dressing. The second-last and last terms in the first line of \eqref{e:H} describe so-called {\em spin-changing collisions} (SCCs), which are nonlinear interactions transforming a pair of 0-bosons into a $+$ and a $-$ boson, or {\em vice versa}. See \cite{LawPuBigelow98,SchulzMSc} for details of the derivation of \eqref{e:H} from \eqref{e:H_general} and the assumptions and approximations made along the way. Equation \eqref{e:H} is the starting point for all results reported in this paper, and we refer to it as the {\em SCC Hamiltonian}.

The way in which the three species of bosonic operators occur in \eqref{e:H} suggests to introduce the operators
\begin{subequations}
\begin{align}
L_-&=\tfrac{1}{2}a_0^{\phantom{\dagger}} a_0^{\phantom{\dagger}},& K_-&=-a_-^{\phantom{\dagger}} a_+^{\phantom{\dagger}},\label{e:L-K-}\\
L_+&=\tfrac{1}{2}a_0^\dagger a_0^\dagger,& K_+&=-a_-^\dagger a_+^\dagger,\label{e:L+K+}\\
L_z&=\tfrac{1}{2}\bigl(a_0^\dagger a_0^{\phantom{\dagger}}+\tfrac{1}{2}\bigr),& K_z&=\tfrac{1}{2}\bigl(a_-^\dagger a_-^{\phantom{\dagger}}+a_+^\dagger a_+^{\phantom{\dagger}} +1\bigr),\label{e:LzKz}
\end{align}
\end{subequations}
and to write the Hamiltonian as
\begin{equation}\label{e:HLK}
H=\left(4\lambda L_z-2\lambda+q\right)\left(2K_z-1\right)-4\lambda\left(L_-K_+ + L_+K_-\right).
\end{equation}
The $L_\kappa$ and $K_\kappa$ operators defined in \eqref{e:L-K-}--\eqref{e:LzKz} are one-mode, respectively two-mode, representations of the SU(1,1) algebra \cite{Novaes04} satisfying
\begin{subequations}
\begin{align}
\left[L_-,L_+\right]&=2L_z,& \left[L_z,L_\pm\right]&=\pm L_\pm,\label{e:commuL}\\
\left[K_-,K_+\right]&=2K_z,& \left[K_z,K_\pm\right]&=\pm K_\pm,\label{e:commuK}
\end{align}
\end{subequations}
a property that will turn out to be beneficial for treating the Hamiltonian by algebraic techniques.

\section{SU(1,1) interferometry\texorpdfstring{\\}{} in the nondepleted regime}
\label{s:nondepleted}
In 1986, Yurke, McCall, and Klauder pointed out that interferometers can be characterised by certain Lie groups \cite{YurkeMcCallKlauder86}. The group SU(2) naturally characterises passive interferometers like Mach--Zehnder and Fabry--Perot devices. In the same paper the authors introduced a class of active interferometers characterised by the group SU(1,1), and they also proposed an optical realization of such a device by means of active elements such as degenerate-parametric amplifiers and four-wave mixers. Strikingly, in contrast to SU(2) interferometers, SU(1,1) interferometers can achieve a phase sensitivity of $\Delta\phi=1/N$, surpassing the standard quantum limit $\Delta\phi=1/\sqrt{N}$, even without the use of entangled input states \cite{YurkeMcCallKlauder86}.

In the Lie-group-theoretic language, the effect of an SU(1,1) interferometer on an input state $\ket{\text{in}}$ is written as
\begin{equation}\label{e:SU11sequence}
\ket{\text{out}}=e^{-i\beta K_x}e^{-i\phi K_z}e^{i\beta K_x}\ket{\text{in}}
\end{equation}
with $K_x=(K_+ + K_-)/2$ and real parameters $\beta$ and $\phi$. The first and third exponentials in \eqref{e:SU11sequence} contain $K_x$ operators which, being defined in terms of products of $a_-$ and $a_+$ operators, facilitate the creation of entangled pairs of $\pm$-bosons. These exponentials constitute the active components of the interferometer and, for the example of an optical SU(1,1) interferometer, may model the effect of four-wave mixers. The second exponential in \eqref{e:SU11sequence} contains $K_z$ which, being a sum of number operators \eqref{e:LzKz}, corresponds to free propagation in the interferometer; see \cite{YurkeMcCallKlauder86} for a detailed explanation of the Lie-group-theoretic description of interferometers. The experimental realization of an SU(1,1) interferometer then hinges on the ability to implement the interferometric sequence \eqref{e:SU11sequence}, i.e., time evolution under the Hamiltonians $-K_x$, $K_z$, and $K_x$ with evolution times $\beta$, $\phi$, and $\beta$, respectively. Here and in the following we use units where $\hbar=1$, which corresponds to measuring evolution times in units of $1/\text{Joule}$.

The Hamiltonian \eqref{e:HLK}, and hence \eqref{e:H}, can be used to approximately implement $K_x$ by requiring the following conditions to hold.
\begin{enumerate}
\renewcommand{\itemsep}{0mm}
\renewcommand{\theenumi}{(\roman{enumi})}
\renewcommand{\labelenumi}{\theenumi}
\item\label{i:first} The number $N$ of bosons is large, $N\gg1$.
\item\label{i:2} Most of the bosons are in the $\kappa=0$ state, $a_0^\dagger a_0^{\phantom{\dagger}}\approx N$.
\item\label{i:last} $q$ and $\lambda$ are chosen such that $q/\lambda\approx1-2N$.
\end{enumerate}
Conditions \ref{i:2} and \ref{i:last} imply that $4\lambda L_z-2\lambda+q\approx0$ and hence the first term on the right-hand side of \eqref{e:HLK} vanishes. Conditions \ref{i:first} and \ref{i:2} imply that $a_0^\dagger a_0^{\phantom{\dagger}}$ is large, which justifies replacing $a_0$ operators by complex numbers \cite{Bogoliubov47,Abrikosov75}, $a_0^{\phantom{\dagger}}=\sqrt{N}=a_0^\dagger$, which implies $L_\mp K_\pm=(N/2)K_\pm$. The resulting Hamiltonian
\begin{equation}\label{e:Hnd}
H_\text{nd}=-4\lambda N K_x
\end{equation}
in general changes the occupation of the $\pm$-boson modes and is one of the building blocks of SU(1,1) interferometric sequence \eqref{e:SU11sequence}. The subscript ``$\text{nd}$'' of the Hamiltonian \eqref{e:Hnd} stands for {\em nondepleted}, a term that refers to the regime where conditions \ref{i:first}--\ref{i:last} hold and hence the occupation number of the $0$-bosons can be considered as infinite in good approximation and is not depleted appreciably when creating $\pm$-boson pairs from $0$-bosons by means of spin-changing collisions. Switching off all interactions, free phase evolution occurs according to
\begin{equation}\label{e:Hphi}
H_\phi=\omega a_-^\dagger a_-^{\phantom{\dagger}} + \omega_0 a_0^\dagger a_0^{\phantom{\dagger}} + \omega a_+^\dagger a_+^{\phantom{\dagger}}.
\end{equation}
Upon shifting $\omega_0$ to zero, \eqref{e:Hphi} becomes, besides an irrelevant constant term, proportional to $K_z$, which realises the second building block of the interferometer \eqref{e:SU11sequence}.

Based on a gas of ultracold Rubidium atoms in a regime that is described by the Hamiltonian \eqref{e:H}, an experimental realization of an atomic SU(1,1) interferometer has been achieved by using samples of $N\approx500$ atoms, preparing an initial state where all atoms are in the 0 hyperfine state, i.e.,
\begin{subequations}
\begin{align}
\braket{\text{in}|a_0^\dagger a_0^{\phantom{\dagger}}|\text{in}}&=N,\label{e:initial0}\\
\braket{\text{in}|a_-^\dagger a_-^{\phantom{\dagger}}|\text{in}}&=0=\braket{\text{in}|a_+^\dagger a_+^{\phantom{\dagger}}|\text{in}},\label{e:initialpm}
\end{align}
\end{subequations}
by tuning $q/\lambda$ to satisfy condition \ref{i:last}, and by restricting the experiment to evolution times that are short enough for the system to remain in the nondepleted regime where $a_0^\dagger a_0^{\phantom{\dagger}}\approx N$ \cite{SchulzMSc,Linnemann_etal16,Linnemann_etal17}. The measured output in these experiments is the $\pm$-mode occupation
\begin{equation}\label{e:eta}
\eta=2K_z-1=a_-^\dagger a_-^{\phantom{\dagger}}+a_+^\dagger a_+^{\phantom{\dagger}}
\end{equation}
at the end of the interferometric sequence.

\section{Phase sensitivity and Fisher information}
\label{s:Fisher}
Interferometers detect relative phase shifts, which can be determined by measuring interferences of the output beams. A key figure of merit for any interferometer is the phase sensitivity quantified by $(\Delta\phi)^2$. For an interfero\-meter where phase shifts of the observable $\eta$ are measured, a direct way to calculate the phase sensitivity is via the error propagation formula \cite{YurkeMcCallKlauder86}
\begin{equation}\label{e:errorpropagation}
(\Delta\phi)^2 = \left(\Delta\eta\right)^2\bigg/\left(\frac{\partial\langle\eta\rangle}{\partial\phi}\right)^2,
\end{equation}
where expectation values $\langle\eta\rangle$ as well as variances $(\Delta\eta)^2=\langle\eta^2\rangle-\langle\eta\rangle^2$ are calculated with respect to the state $\ket{\text{out}}$. Both $\langle\eta\rangle$ and $(\Delta\eta)^2$ acquire $\phi$-dependences through the interferometric sequence \eqref{e:SU11sequence} that generates $\ket{\text{out}}$. The phase sensitivity of the ideal SU(1,1) interferometer \eqref{e:SU11sequence} can be calculated analytically \cite{YurkeMcCallKlauder86}. For an initial state satisfying \eqref{e:initial0} and \eqref{e:initialpm}, this analytical result simplifies to \cite{MarinoCorzoTrejoLett12}
\begin{equation}\label{e:scale}
(\Delta\phi)^2=\frac{1}{1+\cos\phi}\left[\frac{2}{\eta_1(\eta_1+2)}+1-\cos\phi\right],
\end{equation}
where
\begin{equation}
\eta_1=\Braket{\text{in}|e^{-i\beta K_x}\eta e^{i\beta K_x}|\text{in}}=\cosh\beta-1
\end{equation}
is the number of $\pm$-atoms ``inside'' the interferometer, i.e., after evolution under only the rightmost exponential in the interferometric sequence \eqref{e:SU11sequence}. For $\phi=2\pi n$ with  $n\in\NN$, the phase sensitivity in \eqref{e:scale} scales like $1/\eta_1^2$ asymptotically for large $\eta_1$, which surpasses the standard quantum limit $1/\eta_1$. In Refs.~\cite{Linnemann_etal16,Linnemann_etal17}, the phase sensitivity of an SU(1,1) atom interferometer has been determined experimentally via the error propagation formula \eqref{e:errorpropagation} by measuring $\langle\eta\rangle$ and $(\Delta\eta)^2$ over a range of phases $\phi$, confirming a precision beyond the standard quantum limit.

As an alternative to the error propagation formula \eqref{e:errorpropagation}, one can use the Cramer--Rao bound \cite{BengtssonZyczkowski}
\begin{equation}\label{e:sensitivityFisher}
(\Delta\phi)^2\geqslant\frac{1}{F_I(\phi)},
\end{equation}
to estimate the phase sensitivity in terms of the Fisher information \cite{GiovannettiLloydMaccone11}
\begin{equation}\label{e:Fisher}
F_I(\phi):=\sum_\eta \frac{1}{P_\eta(\phi)}\left(\frac{\partial P_\eta(\phi)}{\partial \phi}\right)^2,
\end{equation}
where
\begin{equation}\label{e:Peta}
P_\eta(\phi)=\left\lvert\braket{\eta/2|\text{out}}\right\rvert^2
\end{equation}
is a probability distribution over the occupation numbers $\eta$. $\ket{\eta/2}$ denotes the Fock state made up of an equal number $\eta/2$ of $+$ and $-$ atoms (which is one of the elements of the Fock basis \eqref{e:Fockbasis} defined later).

To determine $F_I$ directly from its definition \eqref{e:Fisher}, the derivative of $P_\eta$ with respect to $\phi$ needs to be computed, which requires knowledge of the functional dependence of $P_\eta$ on $\phi$. Experimentally determining this functional dependence is challenging, but can be circumvented by resorting to the Hellinger distance \cite{SchulzMSc,Strobel_etal14}
\begin{equation}\label{e:Hellinger}
dH^2_{\phi,\phi+\Delta}=\frac{1}{2}\sum_\eta \left(\sqrt{P_\eta(\phi)}-\sqrt{P_\eta(\phi+\Delta)}\right)^2,
\end{equation}
which does not contain derivatives of $P_\eta$. Taylor-expanding $dH^2_{\phi,\phi+\Delta}$ for small $\Delta$, one finds
\begin{equation}\label{e:HellingerFisher}
dH^2_{\phi,\phi+\Delta}=\frac{1}{8}F_I(\phi)\Delta^2 + \mathscr{O}(\Delta^3),
\end{equation}
and this leading-order proportionality permits to infer $F_I$ from measurements of $dH^2_{\phi,\phi+\Delta}$ \cite{SchulzMSc,Strobel_etal14}.

\section{SCC interferometry}
\label{s:SCC}
The theoretical analysis of an SU(1,1) atom interferometer realised by means of the three-species bosonic Hamiltonian \eqref{e:H} reviewed in Sec.~\ref{s:nondepleted} is mostly based on the simplifying assumptions \ref{i:first}--\ref{i:last} made in that section, which give rise to the nondepleted Hamiltonian \eqref{e:Hnd}. It is the main purpose of the present paper to deal with effects beyond this idealised nondepleted Hamiltonian, which are inevitably present in real experiments. For a realistic description of an active atom interferometer, the time evolution under $K_x$ operators in the idealised interferometric sequence \eqref{e:SU11sequence} needs to be replaced by evolution under the full three-species Hamiltonian, resulting in the interferometric sequence
\begin{equation}\label{e:freesequence}
\ket{\text{out}}=e^{-it H(q,\lambda)}e^{-iu H_\phi}e^{it H(q,\lambda)}\ket{\text{in}},
\end{equation}
where the notation $H(q,\lambda)$ highlights a certain choice of the parameters $q$ and $\lambda$ in \eqref{e:H}. We call the parameter $t$ the {\em seeding time}, during which pairs of $\pm$-bosons are produced via spin-changing collisions; and the parameter $u$ the {\em dwell time}, during which free phase evolution takes place. Even the free phase evolution $\exp(-iu H_\phi)$ under the Hamiltonian $H_\phi$ in Eq.~\eqref{e:Hphi} may be seen as an idealization of the actual experimental protocol: The interferometric sequence \eqref{e:freesequence} describes a time evolution under a Hamiltonian that switches instantaneously from $H(q,\lambda)$ to $H_\phi$ and back. In practice it is difficult to change the interaction strength $\lambda$ fast enough to achieve even an approximately instantaneous switch [which would require a change that is much faster than any of the intrinsic timescales of the dynamics under $H(q,\lambda)$ and $H_\phi$]. Instead, in the experimental realization of the interferometer reported in Refs.~\cite{Linnemann_etal16,Linnemann_etal17}, a quasifree evolution is implemented by switching the interacting Hamiltonian \eqref{e:H} to a large value of $q$, such that the interferometric sequence is given by
\begin{equation}\label{e:quasifreesequence}
\ket{\text{out}'}=e^{-it H(q,\lambda)}e^{-iu H(q',\lambda)}e^{it H(q,\lambda)}\ket{\text{in}}
\end{equation}
with $q'\gg q,\lambda$. In this limit, the Hamiltonian \eqref{e:H} is given by
\begin{equation}
H(q',\lambda)=q'\bigl[a_-^\dagger a_-^{\phantom{\dagger}} + a_+^\dagger a_+^{\phantom{\dagger}}+\mathscr{O}(\lambda/q')\bigr]
\end{equation}
to leading order in the small parameter $\lambda/q'$, justifying the claim of a {\em quasifree}\/ phase evolution \footnote{The actual experimental sequence in Refs.~\cite{Linnemann_etal16,Linnemann_etal17} is $\exp[-it H(q,\lambda)] \exp[-iu H(q',\lambda)]\exp[-it H(q,\lambda)]\ket{\text{in}}$, which has the advantage of not requiring a sign inversion of the Hamiltonian and may hence be easier to implement. The main difference of such a protocol is that, unlike in Fig.~\ref{f:sensitivity}, the optimal phase sensitivity is achieved not in the vicinity of $\phi=0$ (or multiples of $2\pi$), but closer to $\phi=\pi$. The Bethe-Ansatz techniques developed in the present paper can be applied to this modified interferometric sequence in just the same way.}.
 
Unlike in the case of the ideal SU(1,1)-interferometer \eqref{e:SU11sequence}, the phase $\phi$ does not feature as a parameter in the interferometric sequence \eqref{e:quasifreesequence}. Instead, $\phi$ is expected to be approximately proportional to the quasifree evolution time $u$, at least for those parameter values for which the device indeed functions as an interferometer. We will come back to this issue, and determine the proportionality constant between $\phi$ and $u$, in Sec.~\ref{s:sensitivityrapidities}.

Effects beyond the ideal SU(1,1) interferometric sequence \eqref{e:SU11sequence} are certainly harder to deal with theoretically. However, we show in the following that the full Hamiltonian \eqref{e:HLK} is amenable, for arbitrary parameter values, to an exact analytic treatment by means of Bethe Ansatz techniques. These methods allow us to calculate phase sensitivities and the Hellinger distance essentially analytically for either of the interferometric protocols \eqref{e:freesequence} or \eqref{e:quasifreesequence}. Since these protocols are based on the Hamiltonian \eqref{e:HLK} that models spin-changing collisions (SCC), we refer to both sequences  \eqref{e:freesequence} and \eqref{e:quasifreesequence} as {\em SCC interferometry}.

\section{Bethe Ansatz solution}
\label{s:Bethe}
The Hamiltonian \eqref{e:HLK} satisfies the conditions of a Richardson--Gaudin model \cite{DukelskyEsebbagSchuck01,DukelskyPittelSierra04,ClaeysPhD} and therefore its exact eigenstates and eigenvalues can be determined by techniques that fall into the broader class of the algebraic Bethe Ansatz. More specifically, our model is an example of a bosonic pairing model of the type analysed in Ref.~\cite{DukelskyEsebbagSchuck01}. We will make use of results from that study, suitably adapted, in what follows. The appendix contains more information on the derivation of these results, and on how the SCC Hamiltonian fits into the general pairing model formalism.

The starting point for obtaining the solution is the observation that the $L_+$ and $K_+$ operators in \eqref{e:L+K+} can be used to span subspaces of the Fock space of the bosonic system,
\begin{equation}\label{e:Hnupm}
\mathscr{F}_{\bm{\nu}}^\pm = \Span \left\{L_+^l K_+^k\ket{\bm{\nu}_\pm}\,\big|\,l,k\in\NN_0\right\},
\end{equation}
where 
\begin{equation}
\ket{\bm{\nu}_\pm} \equiv (a_0^\dagger)^{\nu_0} (a_\pm^\dagger)^{\nu_1}\ket{0}
\end{equation}
and each subspace is labelled by the {\em seniorities}\/ $\nu_0\in\{0,1\}$ and $\nu_1\in\NN_0$ in the multi-index $\bm{\nu}=(\nu_0,\nu_1)$. The seniorities can be interpreted as the numbers of unpaired 0-bosons and $\pm$-bosons, respectively. The subspaces defined in \eqref{e:Hnupm} are closed under the application of the $L_\kappa$ and $K_\kappa$ operators defined in \eqref{e:L-K-}--\eqref{e:LzKz},
\begin{equation}
L_\kappa\ket{\bm{n}}\in\mathscr{F}_{\bm{\nu}}^\pm\;\Longleftrightarrow\;\ket{\bm{n}}\in\mathscr{F}_{\bm{\nu}}^\pm\;\Longleftrightarrow\; K_\kappa\ket{\bm{n}}\in\mathscr{F}_{\bm{\nu}}^\pm.
\end{equation}

We use Richardson's Ansatz \cite{Richardson68}
\begin{equation}\label{e:Ansatz}
\ket{\psi_s}:=\prod_{\alpha=1}^n \left(\frac{L_+}{1-e_{s\alpha}}-\frac{K_+}{1+e_{s\alpha}}\right)\ket{\bm{\nu}_\pm}
\end{equation}
for states with a specified seniority $\bm{\nu}$ and number of pairs $n$. Our aim is to determine the so-called rapidities $e_{s\alpha}$ in such a way that the states \eqref{e:Ansatz} are eigenstates of the Hamiltonian \eqref{e:HLK}. The index $s$ in \eqref{e:Ansatz} labels different eigenstates of $H$, each of which is specified by a different set $\{e_{s\alpha}\}_{\alpha=1,\dotsc,n}$ of rapidities. In Ref.~\cite{DukelskyEsebbagSchuck01} it is shown that if, for all $\alpha=1,\dotsc,n$, the rapidities satisfy the Richardson equation
\begin{equation}\label{e:Richardson}
1+4g\left(\frac{d_0}{1-e_{s\alpha}}-\frac{d_1}{1+e_{s\alpha}}\right)-4g\sum_{\beta\neq\alpha}\frac{1}{e_{s\alpha}-e_{s\beta}} = 0
\end{equation}
with $g=2\lambda/q$, $d_0=(\nu_0+1/2)/2$, and $d_1=(\nu_1+1)/2$, then
\begin{equation}
H\ket{\psi_s}=E_s\ket{\psi_s}
\end{equation}
holds with
\begin{subequations}
\begin{align}
E_s&=2\lambda-q-4\lambda r_{0s}+2(q-2\lambda)r_{1s},\label{e:energy}\\
r_{0s}&=d_0\biggl(1-2gd_1-4g\sum_{\alpha}\frac{1}{1-e_{s\alpha}}\biggr),\\
r_{1s}&=d_1\biggl(1+2gd_0+4g\sum_{\alpha}\frac{1}{1+e_{s\alpha}}\biggr).\label{e:r1}
\end{align}
\end{subequations}
In this way the task of determining the eigenstates and eigenvalues of the Hamiltonian \eqref{e:HLK} is reduced to finding the roots of a set of $n$ coupled nonlinear algebraic equations \eqref{e:Richardson}. These equations have $n+1$ (in general different) sets of roots, each of which corresponds to a different eigenstate of $H$. For the present bosonic case these rapidities are known to be real \cite{DukelskyEsebbagSchuck01,PittelDukelsky03}. The numerical calculation of the roots of \eqref{e:Richardson} is done by a mapping to the zeros of special polynomials \cite{MarquetteLinks12} and computation of these zeros by standard numeric libraries.

\section{Output states in terms of rapidities}
\label{s:out}
Our aim is to use the Bethe Ansatz solution \eqref{e:Ansatz}--\eqref{e:r1} to compute the state $\ket{\text{out}}$ at the end of either of the interferometric sequences \eqref{e:freesequence} or \eqref{e:quasifreesequence}. This output state, in turn, can then be used to calculate expectation values like $\braket{\text{out}|\eta|\text{out}}$, or the probabilities $P_\eta$ in \eqref{e:Peta} that are required for computing the Fisher information \eqref{e:Fisher} or the Hellinger distance \eqref{e:Hellinger}. The three exponentials occurring in \eqref{e:freesequence} or \eqref{e:quasifreesequence} are most easily evaluated in the respective eigenbases of the Hamiltonians. To this aim, we introduce transformations between the relevant bases, which then allow us to evaluate the sequence of three time evolutions successively and write $\ket{\text{out}}$ in terms of transformation matrix elements and phase factors.

Guided by the experimental realizations \cite{Linnemann_etal16,Linnemann_etal17}, we assume the initial state to consist of $2n$ bosons in the 0-hyperfine state,
\begin{equation}\label{e:normin}
\ket{\text{in}}=\frac{1}{\sqrt{2^{-2n}(2n)!}}L_+^n\ket{0}.
\end{equation}
This state is in the seniority sector $\nu_0=0=\nu_1$ of the Hilbert space and, since all evolution operators in \eqref{e:freesequence} and \eqref{e:quasifreesequence} conserve seniority as well as boson number, the system will remain in that sector throughout the interferometric sequence. One distinguished basis is therefore the orthonormalised $2n$-boson Fock basis $\left\{\ket{k}\right\}_{k=0}^n$ with
\begin{equation}\label{e:Fockbasis}
\ket{k}=\frac{1}{\sqrt{N_k}}L_+^{n-k}K_+^k\ket{0},
\end{equation}
where $N_{k}=2^{2(k-n)}[2(n-k)]!(k!)^2$, and it follows that in this basis $\ket{\text{in}}=\ket{0}$.

The eigenstates \eqref{e:Ansatz} of the Hamiltonian \eqref{e:HLK}, which form a second distinguished basis $\left\{\ket{\psi_s}\right\}_{s=0}^n$ for describing the interferometer, have been derived in Sec.~\ref{s:Bethe}. In our selected seniority sector these states can be written by means of a binomial-type expansion as
\begin{equation}\label{e:energybasis}
\ket{\psi_s}=\frac{1}{\sqrt{\mathscr{N}_s}}\sum_{k=0}^n \alpha_{sk}\ L_+^{n-k}K_+^{k}\ket{0},
\end{equation}
where $\mathscr{N}_s=\sum_{k=0}^{n} N_k\left\lvert\alpha_{sk}\right\rvert^2$,
\begin{equation}\label{e:alpha}
\alpha_{sk}:=\sum_{P\in S_n}\prod_{\alpha=1}^k \frac{-1}{1+e_{sP(\alpha)}}\prod_{\beta=k+1}^n \frac{1}{1-e_{sP(\beta)}},
\end{equation}
and $S_n$ denotes the symmetric group consisting of all permutations of a set of $n$ elements, and $P(\alpha)$ denotes the element onto which $\alpha$ is permuted upon application of $P\in S_n$. The energy basis and the Fock basis are linked via
\begin{equation}\label{e:basistrafo}
\ket{\psi_s}=\sum_{k=0}^n c_{sk}\ket{k}
\end{equation}
with
\begin{equation}\label{e:c}
c_{sk}=\alpha_{sk}\sqrt{N_k/\mathscr{N}_s},
\end{equation}
where the elements \eqref{e:c} define an $(n+1)\times(n+1)$ unitary transformation matrix. By means of this transformation, the initial state \eqref{e:normin} can be written in the energy eigenbasis as
\begin{equation}
\ket{\text{in}}=\sum_{s=0}^{n}c_{s0}^*\ket{\psi_s}.
\end{equation}

\begin{figure}
\includegraphics[width=0.8\linewidth]{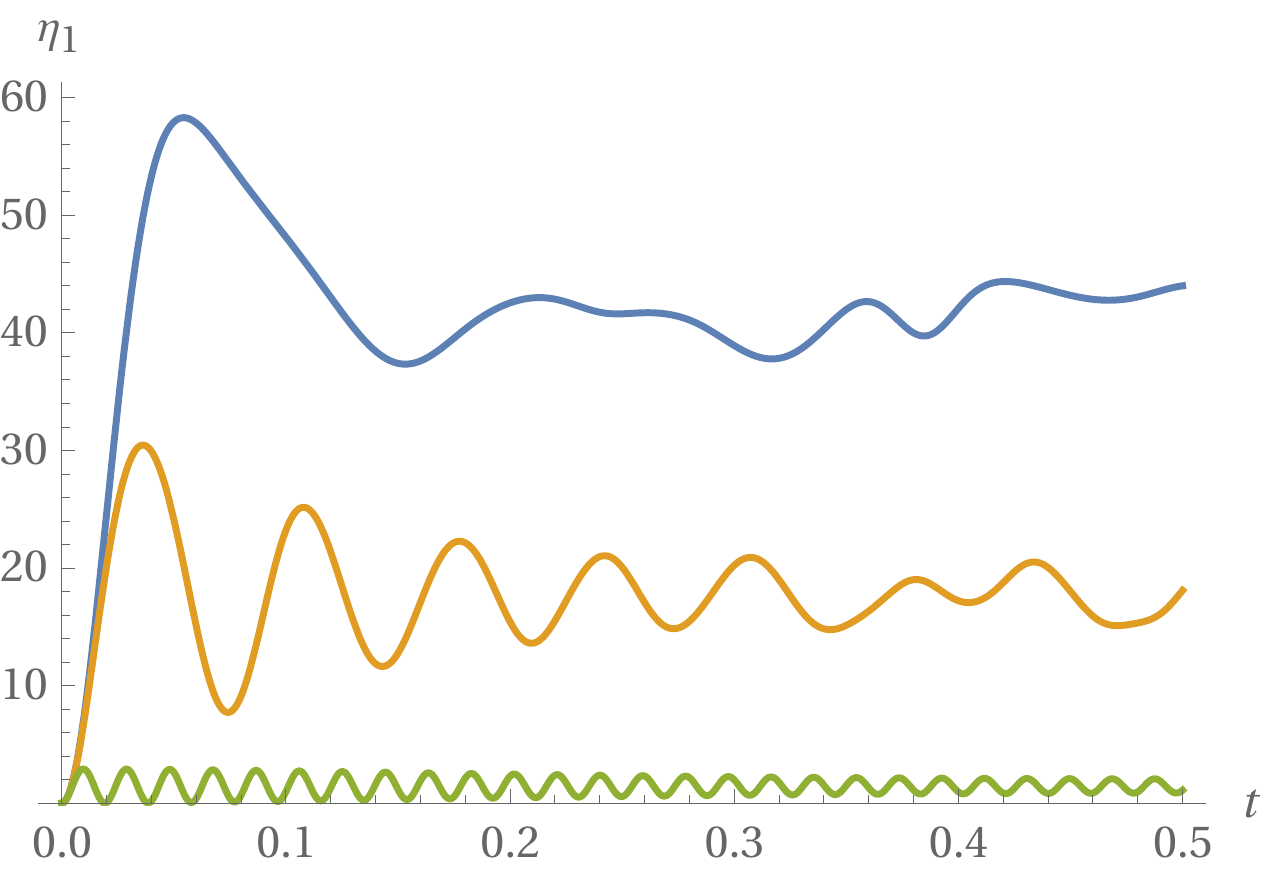}
\caption{\label{f:scc}
The expectation values $\eta_1$ from Eq.~\eqref{e:eta1}, plotted as functions of $t$ for parameter values $N=100$ and $\lambda=1$. The three different curves are, from top to bottom, for $q=4/3$ (blue), $q=6$ (orange), and $q=60$ (green).
}
\end{figure}

As a first application, we calculate the expectation value of the number of $\pm$-bosons produced by the first (rightmost) of the three evolution factors in the interferometric sequences \eqref{e:freesequence} and \eqref{e:quasifreesequence}. Using the basis transformation \eqref{e:basistrafo} and the fact that the Fock states \eqref{e:Fockbasis} are eigenstates of the $\pm$-boson number operator $\eta$ \eqref{e:eta}, one obtains the expression
\begin{align}\label{e:eta1}
\eta_1&=\Braket{\text{in}|e^{-it H}\eta e^{it H}|\text{in}}\nonumber\\
&=2\sum_{s,r,k=0}^{n}c_{r0}^{\phantom{*}}c_{rk}^*c_{sk}^{\phantom{*}}c_{s0}^*e^{i(E_s-E_r)t}k,
\end{align}
where $E_r$ and $E_s$ are eigenenergies \eqref{e:energy} of $H$. Analytical expressions for other expectation values can be calculated in a similar manner. The nontrivial ingredients on the right-hand side of Eq.~\eqref{e:eta1} are the $c$-coefficients \eqref{e:c}, which, via $\alpha_{sk}$ defined in \eqref{e:alpha}, depend on the rapidities $e_{s\alpha}$. The rapidities are determined numerically from the Richardson equations \eqref{e:Richardson} as outlined at the end of Sec.~\ref{s:Bethe}. Figure~\ref{f:scc} shows the dependence of the expectation value $\eta_1$ \eqref{e:eta1} on the duration $t$ of the first ``active'' phase of the interferometric sequences \eqref{e:freesequence} or \eqref{e:quasifreesequence}. The larger $\eta_1$, the more $\pm$-bosons are available ``inside'' the interferometer as a resource of entanglement, which is essential for the surpassing of the standard quantum limit. We observe that, in the regime where $\lambda$ and $q$ are of similar magnitude and for sufficiently long seeding times $t$, a substantial number of $\pm$-bosons is produced, fluctuating roughly around $N/2$. For large values of $q$, the production of $\pm$-bosons is strongly suppressed and $\eta$ oscillates in an approximately sinusoidal fashion around a value much smaller than $N/2$. This confirms, as discussed towards the end of Sec.~\ref{s:Fisher}, that time evolution under the Hamiltonian $H(q,\lambda)$ with $q\gg\lambda$ can be used to approximate free phase evolution, as proposed in the interferometric sequence \eqref{e:quasifreesequence}.

Similar to the derivation of Eq.~\eqref{e:eta1}, the output state at the end of the full interferometric sequence \eqref{e:freesequence} can be evaluated by repeated transformations between the Fock basis \eqref{e:Fockbasis} and the energy basis \eqref{e:energybasis}, yielding
\begin{equation}\label{e:outc}
\ket{\text{out}}=\sum_{q=0}^n x_q(u,t)\ket{\psi_q}
\end{equation}
with
\begin{equation}\label{e:x_coeff}
x_q(u,t)=\sum_{s,r=0}^n c_{qr}^* c_{sr}^{\phantom{*}} c_{s0}^* e^{i(E_s-E_q)t}  e^{-2i[r\omega+(n-r)\omega_0]u}.
\end{equation}

\begin{figure}[t]
\includegraphics[width=0.48\linewidth]{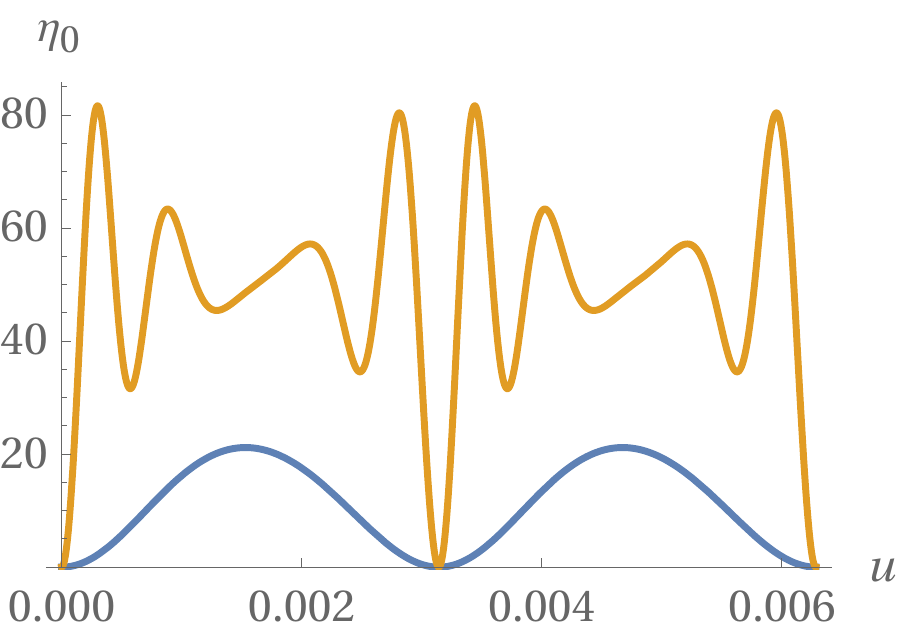}
\includegraphics[width=0.48\linewidth]{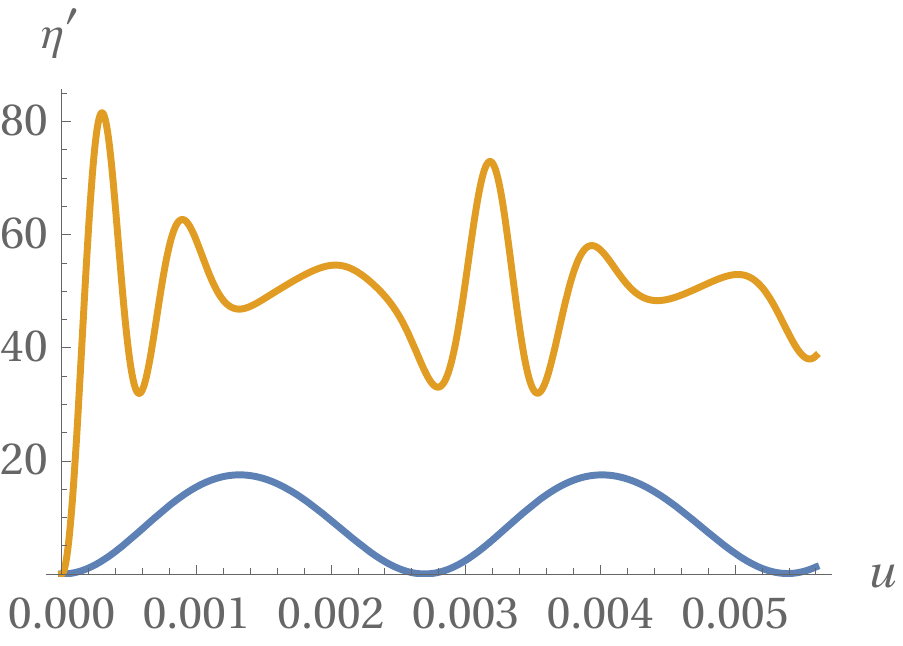}
\caption{\label{f:etaout}
Left: Expectation value $\eta_0\equiv\braket{\text{out}|\eta|\text{out}}$ at the end of the interferometric sequence \eqref{e:freesequence} with free phase evolution, calculated for parameter values $N=100$, $q=4/3$, $\lambda=1$, and $\omega=1000$, and plotted as a function of the dwell time $u$ inside the interferometer. The blue curve uses a seeding time $t=0.006$, which, according to Fig.~\ref{f:scc}, corresponds to a production of approximately $1.5$ pairs of $\pm$-bosons at the end of the first (rightmost) exponential in the sequence \eqref{e:freesequence}. The orange curve is for $t=0.03$, corresponding to the creation of approximately $21$ pairs.
Right: As in the left plot, but showing the expectation value $\eta^\prime\equiv\braket{\text{out}'|\eta|\text{out}'}$ calculated for the interferometric sequence \eqref{e:quasifreesequence} with quasifree phase evolution and $q'=1000$.
}
\end{figure}

For the interferometric sequence \eqref{e:quasifreesequence}, in which free phase evolution is replaced by evolution under $H(q',\lambda)$ with $q'$ large, we transform, in addition to the eigenbasis \eqref{e:energybasis} of $H(q,\lambda)$, also to the eigenbasis $\{\ket{\psi_s^\prime}\}_{s=0}^n$ of $H(q',\lambda)$. Expansion coefficients $\alpha'$ and basis transformation coefficients $c'$ are defined analogous to their non-primed counterparts \eqref{e:alpha} and \eqref{e:c}. With this notation, by repeated transformations between the Fock basis \eqref{e:Fockbasis}, the energy basis \eqref{e:energybasis}, and the primed energy basis, the output state at the end of the interferometric sequence \eqref{e:quasifreesequence} can be written as
\begin{equation}\label{e:outprimec}
\ket{\text{out}'}=\sum_{q=0}^n x_q^\prime(u,t)\ket{\psi_q}
\end{equation}
with
\begin{equation}\label{e:xprime_coeff}
x_q^\prime(u,t)=\sum_{m,p,r,s=0}^n c_{qp}^* c_{mp}^{\prime} c'^*_{mr} c_{sr}^{\phantom{*}} c_{s0}^* e^{i(E_s-E_q)t} e^{-iE_m^\prime u}.
\end{equation}
By expanding $\ket{\psi_q}$ in \eqref{e:outc} or \eqref{e:outprimec} in the Fock basis \eqref{e:Fockbasis}, which is an eigenbasis of $\eta$, we can evaluate the expectation values $\eta_0\equiv\braket{\text{out}|\eta|\text{out}}$ and $\eta^\prime\equiv\braket{\text{out}'|\eta|\text{out}'}$ at the end of the respective interferometric sequences. Figure~\ref{f:etaout} (left) shows, for two choices of the seeding time $t$, $\eta_0$ as a function of the dwell time $u$. In the case of a short seeding time $t=0.006$ (blue) we observe clear interference fringes with an oscillation period of $0.003$. As in the case of the ideal SU(1,1) interferometer (Eq. (9.28) in Ref.~\cite{YurkeMcCallKlauder86}) the fringes are approximately sinusoidal, which makes this regime particularly suitable for interferometry. For longer seeding time $t=0.03$ (orange) the same fundamental period of $0.003$ is observed, but with higher frequency contributions superimposed. Figure~\ref{f:etaout} (right) shows similar data, but for the expectation value $\eta^\prime$ calculated for the interferometric sequence \eqref{e:quasifreesequence} with quasifree phase evolution. In this case, the strict periodicity is spoiled, which is particularly evident for the example with the larger seeding time $t$ (orange).\\

\section{Phase sensitivity in terms of rapidities}
\label{s:sensitivityrapidities}
To calculate the phase sensitivities of the interferometric sequences \eqref{e:freesequence} or \eqref{e:quasifreesequence}, we relate the dwell time $u$ to the phase $\phi$. This is achieved by numerically determining $\eta_0$ or $\eta^\prime$ as a function of $u$. If, as for the parameter values $\lambda$, $q$, and $q^\prime$ in Fig.~\ref{f:etaout} (right), a roughly periodic dependence on $u$ is observed, then the angular frequency $\Omega$ of the oscillatory behaviour can be read off, and we identify $\phi=\Omega u$. The output states $\ket{\text{out}}$ \eqref{e:outc} and $\ket{\text{out}'}$ \eqref{e:outprimec} depend on the dwell time $u$, and hence on the interferometric phase $\phi$, only through the exponentials in the coefficients \eqref{e:x_coeff} and \eqref{e:xprime_coeff}, respectively. Taking derivatives with respect to $\phi$, as required for the calculation of the phase sensitivity \eqref{e:errorpropagation}, can therefore be done analytically. Using the output state \eqref{e:freesequence} with coefficients \eqref{e:x_coeff} to calculate the expectation value $\braket{\text{out}|\eta|\text{out}}$ and then taking its derivative with respect to $\phi$, the phase sensitivity $(\Delta\phi)^2$ \eqref{e:errorpropagation} can be expressed in terms of the rapidities $e_{s\alpha}$,
\begin{widetext}
\begin{equation}\label{e:DeltaPhi}
(\Delta\phi)^2=\frac{\displaystyle \sum_{p,m,k=0}^nx^*_px_mc_{mk}^{\phantom{*}}c_{pk}^*k^2-\Biggl(\sum_{p,m,k=0}^nx^*_px_mc_{mk}^{\phantom{*}}c_{pk}^*k\Biggr)^2}{\displaystyle \Biggl(\sum_{p,m,k=0}^n(\partial_\phi x^*_p)x_mc_{mk}^{\phantom{*}}c_{pk}^*k+\sum_{p,m,k=0}^nx^*_p(\partial_\phi x_m)c_{mk}^{\phantom{*}}c_{pk}^*k\Biggr)^2}.
\end{equation}
\end{widetext}
For the interferometer with quasifree time evolution \eqref{e:quasifreesequence} the same formula holds with coefficients $x$ replaced by $x'$.

\begin{figure}
\includegraphics[width=0.49\linewidth]{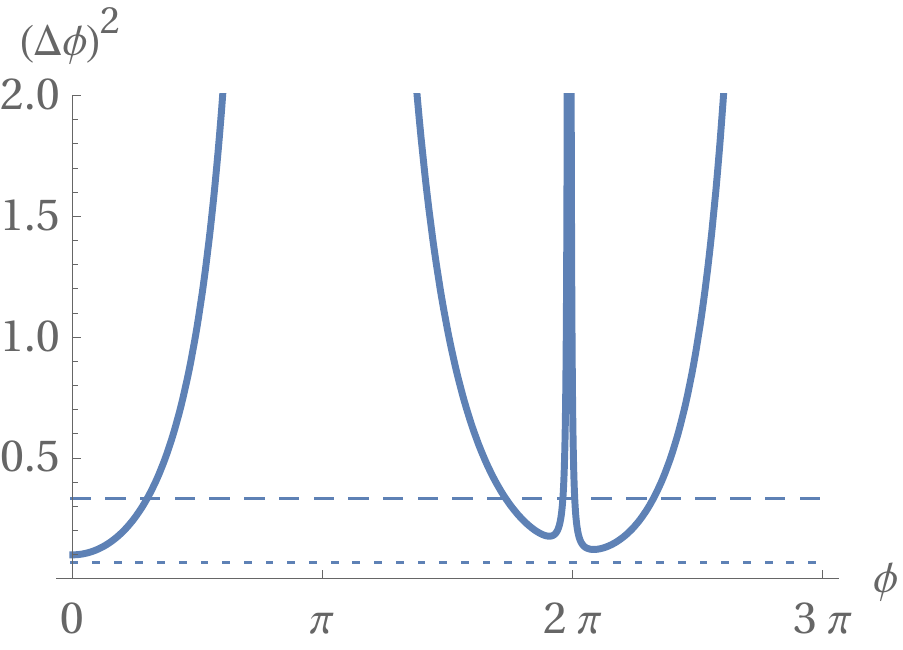}
\includegraphics[width=0.49\linewidth]{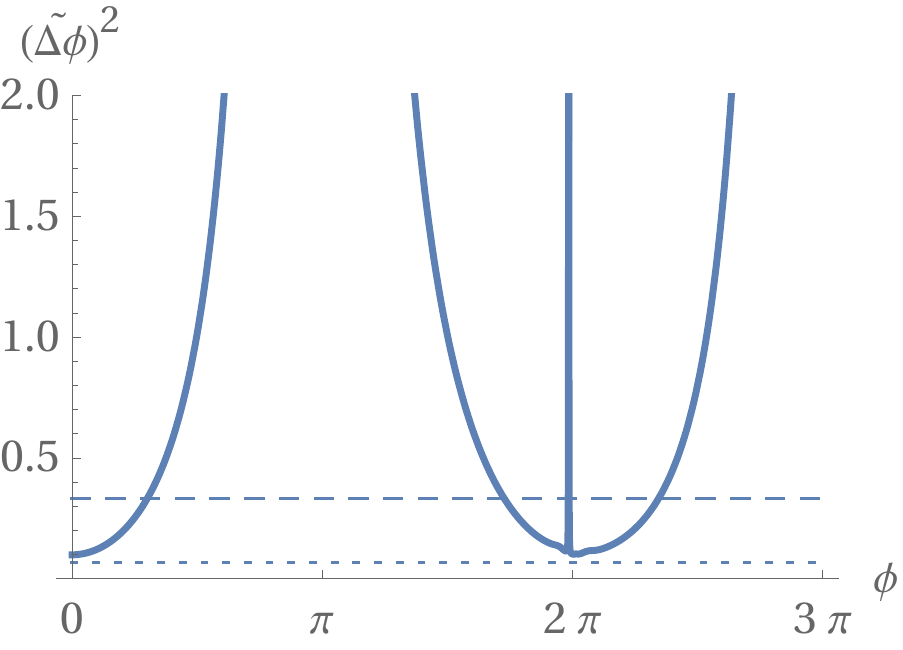}
\caption{\label{f:sensitivity}
Sensitivity of the quasifree interferometer \eqref{e:quasifreesequence} as a function of the phase. For the device to function as an interferometer, we choose parameter values such that $\eta'$ shows sinusoidal oscillations to a good approximation, as for the blue line in Fig.~\ref{f:etaout} (right), but unlike for the yellow line in that figure. This is achieved by setting $N=100$, $\lambda=1$, $q=4/3$, $q'=1000$, and a seeding time $t=0.006$, which, according to the blue curve in Fig.~\ref{f:etaout}, results in oscillations with an angular frequency $\Omega\approx2307$. This frequency relates the phase $\phi=\Omega u$ to the dwell time $u$. The phase sensitivity is calculated by means of the error propagation formula \eqref{e:DeltaPhi} in the left plot and via the Hellinger distance \eqref{e:tildeDeltaphi} with $\Delta=10^{-5}$ in the right plot. While the plots show minor differences, the results are generally in good agreement. In both cases the phase sensitivities fall significantly below the standard quantum limit $1/\eta_1\approx1/3$ (dashed lines), and closely approach the Heisenberg limit $1/[\eta_1(\eta_1+2)]\approx1/15$ (dotted lines). Results for the free interferometer \eqref{e:freesequence} are similar (not shown).}
\end{figure}

Figure~\ref{f:sensitivity} (left) shows the sensitivity $(\Delta\phi)^2$ of the quasifree interferometer \eqref{e:quasifreesequence} as a function of the phase $\phi$. The sensitivity exhibits pronounced minima when the phase is around multiples of $2\pi$, affirming that this is where the interferometer, like its ideal SU(1,1) counterpart, performs at its most precise. We will choose $\phi=0$ for all numerical explorations from here on. As illustrated in the plot, the minimum value of $(\Delta\phi)^2$ is well below the standard quantum limit $1/\eta_1$ and approaches fairly closely, but does not quite reach, the sensitivity $1/[\eta_1(\eta_1+2)]$ of the ideal SU(1,1) interferometer \eqref{e:scale}. Similar behaviour is found for the interferometric sequence \eqref{e:freesequence} with free phase evolution (not shown).

\section{Hellinger distance in terms of rapidities}
\label{s:Fisherrapidities}

As an alternative method for estimating the phase sensitivity, the Fisher information \eqref{e:Fisher} or the Hellinger distance \eqref{e:Hellinger} can be expressed in terms of the rapidities $e_{s\alpha}$, which in turn yield estimates of $(\Delta\phi)^2$ via Eqs.~\eqref{e:sensitivityFisher} and \eqref{e:HellingerFisher}. We focus here on the Hellinger distance, as it is readily accessible in the Rubidium experiments reported in Refs.\ \cite{Strobel_etal14,SchulzMSc}.

To compute the probabilities $P_\eta(\phi)=\left\lvert\braket{\eta/2|\text{out}}\right\rvert^2$ in the definition \eqref{e:Hellinger} of the Hellinger distance, we write
\begin{multline}
\braket{\eta/2|\text{out}} = \sum_{q=0}^n \Braket{\eta/2|x_q|\psi_q}\\
= \sum_{p,q=0}^n x_q c_{qp}\braket{\eta/2|p} = \sum_{q=0}^n x_q c_{q,\eta/2},
\end{multline}
where we have used Eqs.~\eqref{e:outc} and \eqref{e:basistrafo}. Calculating the modulus squared of this result and plugging it into \eqref{e:Hellinger}, one obtains an expression for the Hellinger distance in terms of the rapidities, which, while being lengthy, is fairly straightforward to evaluate numerically. Based on this computation and making use of Eqs.~\eqref{e:sensitivityFisher} and \eqref{e:HellingerFisher}, we define
\begin{equation}\label{e:tildeDeltaphi}
(\widetilde{\Delta\phi})^2 :=\frac{\Delta^2}{8\,dH_{\phi,\phi+\Delta}^2}
\end{equation}
as a proxy for the phase sensitivity $(\Delta\phi)^2$. In the following we investigate the dependence of $(\widetilde{\Delta\phi})^2$ on parameters in the Hamiltonian \eqref{e:H} and in the interferometric sequences \eqref{e:freesequence} or \eqref{e:quasifreesequence}, with the aim of singling out the parameter regime of optimal performance of the interferometer. We checked that the numerical results reported in this section are insensitive to the choice of the parameter $\Delta$ in the definition of the Hellinger distance \eqref{e:Hellinger}, as long as it is much smaller than $2\pi$. Figure~\ref{f:sensitivity} (right) shows the sensitivity $(\widetilde{\Delta\phi})^2$ as a function of the phase $\phi$. The plot uses the same parameter values as for $(\Delta\phi)^2$ in Fig.~\ref{f:sensitivity} (left). While the plots show minor differences, the results are generally in good agreement, confirming that $(\widetilde{\Delta\phi})^2$ is a valid proxy for the phase sensitivity $(\Delta\phi)^2$, with noticeable differences occurring only in the vicinity of the divergences at multiples of $2\pi$, caused by extremely small numerators and denominators on the right-hand side of Eq.~\eqref{e:errorpropagation} that amplify numerical inaccuracies.

\begin{figure}
\includegraphics[width=0.9\linewidth]{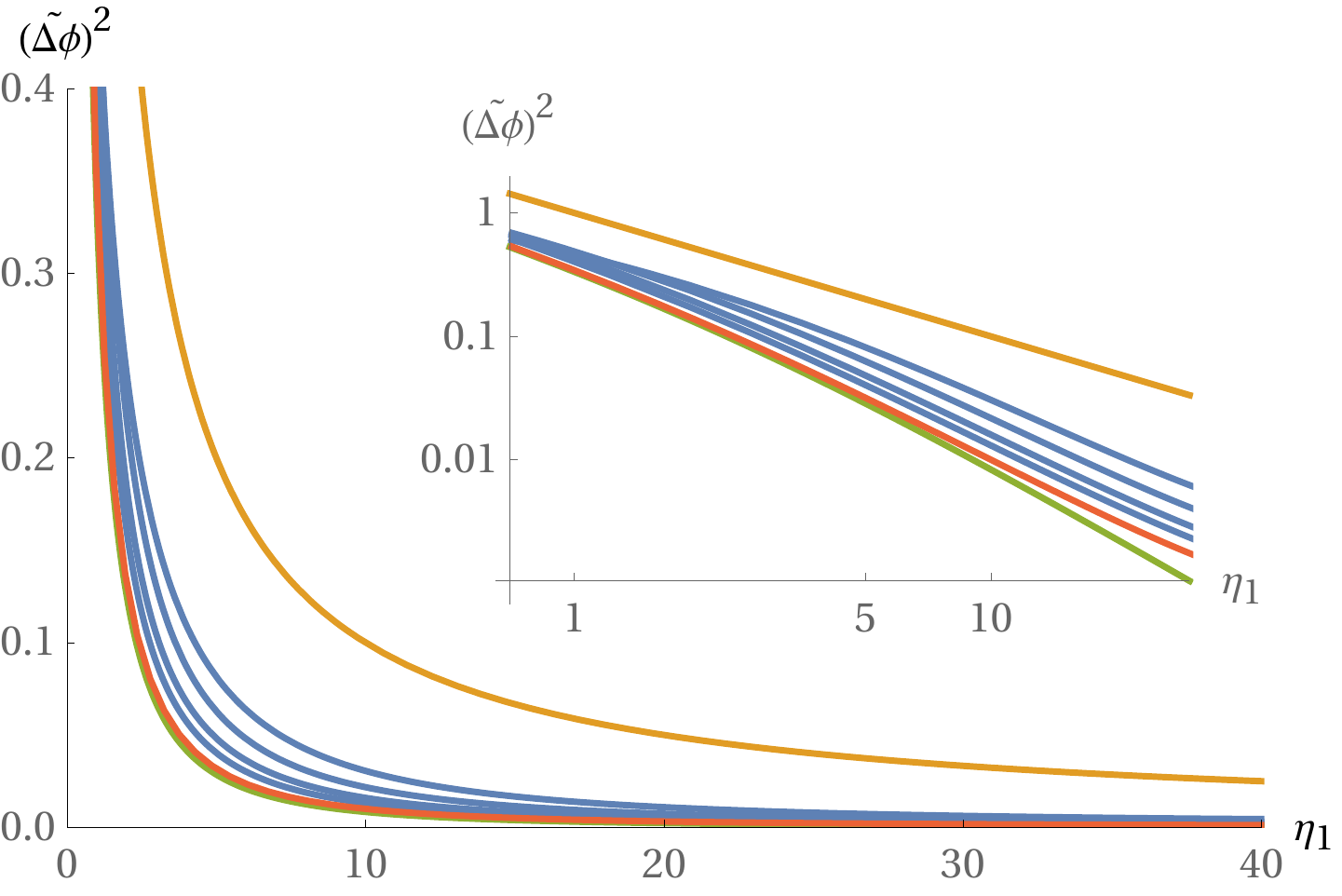}
\caption{\label{f:Fisher_eta1}
Phase sensitivity $(\widetilde{\Delta\phi})^2$ evaluated at its minimum at $\phi=0$, plotted as a function of $\eta_1$ \eqref{e:eta1} for various interferometric sequences. The red line shows $(\widetilde{\Delta\phi})^2$ for the interferometric sequence \eqref{e:freesequence} with free phase evolution for parameter values $N=100$, $\lambda=1$, and $q=4/3$. The blue lines are obtained with the same parameter values, but using the interferometric sequence \eqref{e:quasifreesequence} with quasifree phase evolution and parameter values $q'=125$, 250, 500, and 1000 (top to bottom). All those curves are well below the standard quantum limit $1/\eta_1$ (orange). The phase sensitivity of the free interferometer (red) is very close to the Heisenberg limit $1/[\eta_1(\eta_1+2)]$ (green) of the ideal SU(1,1) interferometer \eqref{e:scale}, and the quasifree interferometer (blue) approaches that limit with increasing $q'$. The inset shows the same data on a logarithmic scale.}
\end{figure}

The main interest in active interferometers, like the ideal SU(1,1) interferometer \eqref{e:SU11sequence} or the SCC interferometers \eqref{e:freesequence} and \eqref{e:quasifreesequence}, lies in their phase sensitivity having the potential to surpass the standard quantum limit, and potentially approach the Heisenberg limit, without the need for entangled input states $\ket{\text{in}}$. Both the standard quantum limit $\sim1/\eta_1$ and the Heisenberg limit $\sim1/\eta_1^2$ are expressed in terms of the number $\eta_1$ of seeded $\pm$-bosons after the first (rightmost) exponential in the interferometric sequences \eqref{e:freesequence} or \eqref{e:quasifreesequence}. To assess the influence of the seeding on the performance of the interferometer, we show in Fig.~\ref{f:Fisher_eta1} the phase sensitivity $(\widetilde{\Delta\phi})^2$ as a function of $\eta_1$ for the free interferometric sequence \eqref{e:freesequence} (red) and the quasifree interferometric sequence \eqref{e:quasifreesequence} (blue). In both cases the sensitivities decay monotonically with $\eta_1$ and behave qualitatively similar to, but are slightly larger than, those of the ideal SU(1,1) interferometer (green line in Fig.~\ref{f:Fisher_eta1}). When operating the interferometer at short seeding times $t$, which results in small values of $\eta_1$, we find, as expected, a phase sensitivity very close to that of the ideal SU(1,1) interferometer. For larger values of $\eta_1$, deviations from the ideal case become visible (see insert of Fig.~\ref{f:Fisher_eta1}), but $(\widetilde{\Delta\phi})^2$ remains well below the standard quantum limit (orange) and decays faster than $1/\eta_1$ asymptotically for large $\eta_1$.

\section{Conclusions}
\label{s:conclusions}
We have theoretically analysed the performance of an active atomic interferometer based on spin-changing collisions (SCCs) in a three-species Bose--Einstein condensate by making use of Bethe Ansatz techniques. Based on the so-called rapidities, which are the solutions of a set of coupled algebraic equations \eqref{e:Richardson}, exact eigenstates and eigenvalues of the Hamiltonian \eqref{e:H} modelling the spin-changing collisions were obtained. These results, in turn, were used to express interferometic quantities, like the phase sensitivity \eqref{e:DeltaPhi} or the related Hellinger distance, in terms of the Bethe rapidities. While the Bethe-Ansatz solution does not necessarily give access to larger system sizes than a straightforward exact diagonalisation of the Hamiltonian, it permits to analytically perform derivatives or similar operations, which may significantly improve numerical accuracies. We use the Bethe-Ansatz solutions to calculate expectation values in the interferometer's output as well as the corresponding phase sensitivities, either directly or via the experimentally more accessible Hellinger distance, which allow us to assess the interferometer's performance.

We studied two versions of the SCC interferometer, one with free phase evolution \eqref{e:freesequence} inside the interferometer, the other one with quasifree phase evolution \eqref{e:quasifreesequence}, which is easier to implement in the existing experimental realisations of an active atomic interferometer \cite{Linnemann_etal16,Linnemann_etal17}. While quasifree evolution spoils the periodicity of the interferometric fringes and operates at slightly inferior phase sensitivity compared to the case of free phase evolution, our results clearly indicate that the SCC interferometer with quasifree phase evolution can successfully function with a phase sensitivity well below the standard quantum limit and, for suitable parameter values, close to the Heisenberg limit accessible by the ideal SU(1,1) interferometer proposed by Yurke, McCall, and Klauder \cite{YurkeMcCallKlauder86}.

While we exploited integrability of the SCC Hamiltonian in order to elegantly and efficiently calculate quantities of interest by expressing them in terms of the Bethe rapidities, integrability does not, to our understanding, affect the performance characteristics of the SCC interferometer. However, the techniques developed in the present paper are general and may potentially be applied to systems governed by the SCC Hamiltonian \eqref{e:H} for applications other than interferometry, either in or out of equilibrium. For example, the equilibration dynamics after a sudden quench of the microwave dressing parameter $q$ in a three-species Bose--Einstein condensate is expected to be strongly affected by integrability \cite{Lamacraft07,DagWangDuan18}, and the Bethe-Ansatz techniques developed in this paper may be brought to use in this context. Neither are applications of this type restricted to the $^{87}$Rb experiments discussed earlier in this paper, but they may also be extended to other alkali-based experiments like $^{23}$Na \cite{Stenger_etal98} and potentially $^7$Li that have three-fold degenerate groundstate manifolds. Moreover, algebraic Bethe Ansatz solutions similar to those employed in the present paper can be constructed for systems consisting of more than three boson species \cite{DukelskyEsebbagSchuck01}, which opens the door for extensions of our methods to atomic species with higher than three-fold degeneracies \cite{Schmaljohann_etal04}.

The numerical evaluations of the Bethe rapidities, or quantities derived from them, reported in this paper are for moderate boson numbers of $N=100$. This particle number can be reached, and exceeded, on a regular desktop computer at the time of writing. Besides the polynomial mapping we used for the calculation of the Bethe rapidities, other numerical approaches have been reported in the literature, and also more efficient methods for the computation of overlaps of Bethe eigenstates are known \cite{FaribaultCalabreseCaux09}. Here, we did not make a concerted effort to reach larger sizes by following any of these and instead opted to focus on conceptual aspects, but we expect that at least an order of magnitude in system size can be gained with a bit of effort. To reach even larger system sizes, and possibly even perform analytical calculations in the large-$N$ limit, the SCC Hamiltonian \eqref{e:HLK} expressed in terms of the generators of the group SU(1,1)$\otimes$SU(1,1) constitutes a promising starting point for Holstein--Primakoff expansions \cite{HolsteinPrimakoff40} or other analytic approaches.

{\em Note added:} When adding the finishing touches to the paper we became aware of the recent preprint Ref.~\cite{Grun_etal} that uses integrability for the analysis of a passive atom interferometer.

\acknowledgments
In an early phase of the project the authors have benefited from helpful discussions with Markus Oberthaler, Frederik Scholtz, and Helmut Strobel.

\appendix*
\section{The SCC Hamiltonian as a solvable pairing model}
\label{s:Richardson}
Here we provide a brief account of the origin of the Richardson equation in \eqref{e:Richardson}, and of the expressions for the eigenvalues of the SCC Hamiltonian \eqref{e:HLK} in Eqs.~\eqref{e:energy}--\eqref{e:r1}. The key observation is that the SCC Hamiltonian falls within the class of solvable models studied in Ref.~\cite{DukelskyEsebbagSchuck01}. In the language of that paper we are dealing with a particular two-level bosonic pairing model. We will label the levels using $l\in\{0,1\}$, where $l=0$ corresponds to the $0$-boson mode, and $l=1$ to the $\pm$-boson modes. The degeneracies of the levels are denoted by $\Omega_0=1$ and $\Omega_1=2$. The $L_\pm$ and $K_\pm$ operators in \eqref{e:L-K-} and \eqref{e:L+K+} create and destroy pairs of bosons in the $l=0$ and $l=1$ levels respectively. Following Ref.~\cite{DukelskyEsebbagSchuck01} we introduce the operators
\begin{subequations}
\begin{align}
R_0 &= L_z+g\left[X_{01}(L_+ K_- + L_- K_+) - 2 Y_{01} L_z K_z\right]\\
R_1 &= K_z+g\left[X_{10}(L_+ K_- + L_- K_+) -\! 2 Y_{10} L_z K_z\right]
\end{align}
\end{subequations}
where $g$, $X_{ll'}$, and $Y_{ll'}$ are scalar parameters. The choice of these parameters is dictated by two requirements, namely that $[R_0,R_1]=0$ and that the SCC Hamiltonian \eqref{e:HLK} can be expressed as a function of these two commuting operators. The former condition is met by setting 
\begin{equation}
	X_{01}=Y_{01}=-X_{10}=-Y_{10}=\frac{1}{\eta_0-\eta_1}
\end{equation}
with $\eta_0$ and $\eta_1$ arbitrary unequal real numbers. This yields the so-called rational model of Ref.~\cite{DukelskyEsebbagSchuck01}. The second requirement is satisfied by choosing
\begin{align}
	g&=2\lambda/q,& \eta_0&=-\eta_1=1/2,
\end{align}
which allows the Hamiltonian \eqref{e:HLK} to be expressed as 
\begin{equation}
H=2\lambda-q-4\lambda R_0+2(q-2\lambda)R_1.
\end{equation}
This implies that $R_0$ and $R_1$ are conserved charges of $H$, and so the eigenstates of $H$ are the simultaneous eigenstates of $R_0$ and $R_1$,
\begin{align}\label{eq:R0R1eigenstates}
R_0\ket{\psi_s}&=r_{0s}\ket{\psi_s},& R_1\ket{\psi_s}&=r_{1s}\ket{\psi_s}.
\end{align}
Here $s$ labels the various eigenstates. Richardson's Ansatz \cite{Richardson68} provides an explicit form for these states as
\begin{equation}
\ket{\psi_s}:=\prod_{\alpha=1}^n \left(\frac{L_+}{2\eta_0-e_{s\alpha}}+\frac{K_+}{2\eta_1-e_{s\alpha}}\right)\ket{\bm{\nu}_\pm},
\end{equation}
which is the origin of Eq.~\eqref{e:Ansatz} in the text. The rapidities $e_{s\alpha}$ are determined by enforcing \eqref{eq:R0R1eigenstates} above. This yields the 
Richardson equations
\begin{equation}
1+4g\sum_l\frac{d_l}{2\eta_l-e_{s\alpha}}-4g\sum_{\beta\neq\alpha}\frac{1}{e_{s\alpha}-e_{s\beta}} = 0
\end{equation}
with $d_l=\left(\nu_l+\Omega_l/2\right)/2$. The eigenvalues of $R_{l\in\{0,1\}}$ are now given in terms of the rapidities as
\begin{equation}
	r_{ls}=d_l\left(1-2g\sum_{l'(\neq l)}\frac{d_{l'}}{\eta_l-\eta_{l'}}-4g\sum_\alpha\frac{1}{2\eta_l-e_{s\alpha}}\right).
\end{equation}
Inserting $\eta_0=-\eta_1=1/2$ into these expressions yields Eqs.~\eqref{e:Richardson} and \eqref{e:energy}--\eqref{e:r1}.

\bibliography{../../../../MK}

\end{document}